\documentclass[prb,aps,twocolumn]{revtex4}
\usepackage{amsmath,bm,epsfig,citesort}

\newcommand{\beq}{\begin{equation}}
\newcommand{\eeq}{\end{equation}}
\newcommand{\bea}{\begin{eqnarray}}
\newcommand{\eea}{\end{eqnarray}}
\newcommand{\bml}{\begin{mathletters}}
\newcommand{\eml}{\end{mathletters}}

\newcommand{\degree}{$^\circ$}

\begin{document}
\draft

\title{\bf Origin of Scaling Behavior of Protein Packing Density: A
Sequential Monte Carlo Study of Compact Long Chain Polymers}

\author{Jinfeng Zhang$^1$, Rong Chen$^{1,2}$, 
Chao Tang$^3$, and 
Jie Liang$^{1,}$\footnote{Corresponding author.  Phone: (312)355--1789, fax: (312)996--5921, email: {\tt jliang@uic.edu}
}
}
\affiliation{
$^1$Departments of Bioengineering and
$^2$Information \& Decision Science\\
 University of Illinois at Chicago, 
845 S.\ Morgan St, Chicago, IL 60607 \vspace*{.1in}\\
$^3$NEC Research Institute, 4 Independence Way, Princeton, NJ 08540
}
\date{\today}

\begin{abstract}
{{\it Single domain proteins are thought to be tightly packed.  The
introduction of voids by mutations is often regarded as destabilizing.
In this study we show that packing density for single domain proteins
decreases with chain length.  We find that the radius of gyration
provides poor description of protein packing but the alpha contact
number we introduce here characterize proteins well.  We further
demonstrate that protein-like scaling relationship between packing
density and chain length is observed in off-lattice self-avoiding
walks.  A key problem in studying compact chain polymer is the
attrition problem: It is difficult to generate independent samples of
compact long self-avoiding walks.  We develop an algorithm based on
the framework of sequential Monte Carlo and succeed in generating
populations of compact long chain off-lattice polymers up to length
$N=2,000$.  Results based on analysis of these chain polymers suggest
that maintaining high packing density is only characteristic of short
chain proteins. We found that the scaling behavior of packing density
with chain length of proteins is a generic feature of random polymers
satisfying loose constraint in compactness.  We conclude that proteins
are not optimized by evolution to eliminate packing voids.
\vspace*{.2in}
}\\

}

\end{abstract}

\maketitle
\newpage
\narrowtext

\section{Introduction}
Geometric considerations have lead to important insights about protein
structures
\cite{Chothia81,Chothia90,Maritan00_Nature,Banavar02_Proteins,Bagci02_JCP}. Voids
are simple geometric features that represent packing defects inside
protein structures.  For multisubunit proteins such as GroEL and
potassium channel, voids or tunnels of large size are formed by the
spatial arrangement of multiple subunits, and are essential for the
biological functions of these proteins
\cite{Sigler98_ARB,Doyle98_Science}.  In this study, we focus on voids
formed due to packing defects that are not directly involved in
protein function.  For this purpose, we choose to study only
structures of single domain proteins.  Although these proteins are
well known to be compact \cite{Richards94_QRB}, and their interior is
frequently thought to be solid-like
\cite{Hermans61_JACS,Richards97_CMLS}, recent calculations showed that
there are also numerous voids buried in the protein interior
\cite{LiangDill01_BJ}.  The importance of tight packing in single
chain protein is widely appreciated: packing is thought to be
important for protein stability
\cite{Eriksson92_Science,Privalov96,Dahiyat97_Science}, for kinetic
nucleation of protein folding \cite{Ptitsyn98_JMB,PtitsynTing99_JMB},
and for successful design of novel proteins following a predefined
backbone \cite{Dahiyat97_Science}. The conservation of amino acid
residues during evolution may also be correlated with tightly packed
sites \cite{Ptitsyn98_JMB,PtitsynTing99_JMB,MirnyShahk01_ARBBS}.  In
contrast, the potential roles of voids in affecting protein stability
and in influencing tolerance to mutations and designability of
proteins \cite{Li96_Science,Melin99_JCP} are not well understood.

An important parameter describing packing is the packing density
$p_d$, which is a quantitative measure of the voids and was first
introduced to study proteins by structural biologists.  This concept
has been widely used in protein chemistry
\cite{Richards94_QRB,Privalov96}.  The scaling relationship of $p_d$
and chain length $N$ was first studied in reference
\cite{LiangDill01_BJ}.  $p_d$ can be thought of as the physical volume
$v_{vdw}$ occupied by the union of van der wall atoms, divided by the
volume of an envelope $v_{env}$ that tightly wraps around the body of
atoms: $p_d \equiv v_{vdw}/v_{env}$ \cite{LiangDill01_BJ}. Voids
contained within the molecule will not be part of the van der Waals
volume $v_{vdw}$, but will be included in $v_{env}$.  Using geometric
algorithms, $v_{vdw}, v_{env}$ and $p_d$ can be readily computed for
protein structures in the Protein Data Bank
\cite{Liang98a_Proteins,Liang98b_Proteins}.

In this work, we further study the scaling behavior of packing density
$p_d$ with chain length of single domain proteins and explore the
determinants of the observed scaling behavior.  We seek to answer the
following questions: Is the scaling behavior of $p_d$ unique to
proteins?  Are proteins optimized during evolution to eliminate
packing voids?  We introduce two new packing parameters $n_\alpha$
(the alpha contact number) and $z_\alpha$ (the alpha coordination
number).  We show that $n_\alpha$ characterizes protein packing very
well with a linear scaling relationship with the chain length, and
that a widely used parameter, the radius of gyration $R_g$,
characterizes protein packing poorly.  To overcome the attrition
problem of low success rate in generating compact long chain polymers,
we develop an algorithm based on sequential Monte Carlo importance
sampling and succeed in obtaining thousands of very compact long chain
off-lattice polymers up to $N=2,000$.  We demonstrate that the scaling
behavior of $p_d$ for proteins can be qualitatively reproduced by
randomly generated polymers with rudimentary constraints of
$n_\alpha$.  Our simulation studies lead us to conclude that proteins
are not optimized to eliminate voids during evolution.  Rather, voids
in proteins are a generic feature of random polymers with a
``reasonable'' (as measure by $z_\alpha$) compactness.

The  paper is organized as follows.  In the Method section, we
first describe briefly how $p_d$ and $n_\alpha$ are computed from
the dual simplicial complex of protein structure, and introduce an
off-lattice discrete model for generating random polymer
conformations.  We next describe the sequential Monte Carlo
importance sampling and resampling techniques that allow us to
generate adequate samples satisfying various criteria of
$n_\alpha$.  In the Results section, we begin with the
characterization of void properties of proteins by both $p_d$ and
$R_g$. We then show the linear scaling behavior of $n_\alpha$
found in proteins.  The scaling behavior of $p_d$ of random
polymers generated by sequential Monte Carlo with chain length is
discussed later.  We conclude with summary and discussion of our
results.

\section{Methods}

{\it Protein Data.}
To avoid complications of multichain and multidomain proteins, we
examine the packing density of proteins of single domain proteins.  We
collect proteins from the {\sc Pdbselect} database \cite{Hobohm94_PS}
that contains only one domain, as defined as single chains in the {\sc
Scop} database with one numerical label \cite{SCOP}.

\begin{figure}[t]
  \centerline{\epsfig{figure=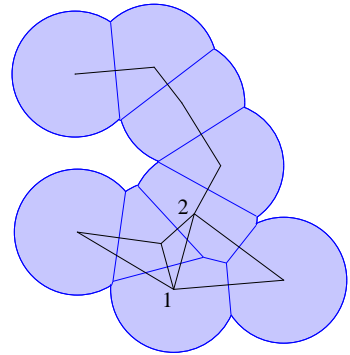,width=3.in}}
\caption{ \sf
The alpha contacts in a toy
molecule.  In this molecule, both atom 1 and atom 2 have 4 alpha
contacts.  The number of atoms $n=9$, the number of alpha contacts is
$n_\alpha = 22$ (twice the number of edges), and the alpha coordination
number $z_\alpha = n_\alpha/n \approx 2.4$.
 }
\label{Fig:AlphaContact}
\end{figure}

{\it Dual Simplicial Complex, Alpha Coordination Number, and Packing
Density.}  We use alpha shape to characterize the geometry of protein
structure.  Alpha shape has been successfully applied to study a
number of problems in proteins, including void measurement, binding
site characterization, protein packing, electrostatic calculations,
and protein hydrations
\cite{Edels95_Hawaii,Liang98a_Proteins,Liang98b_Proteins,Peters96_JMB,Liang98_PS,LiangDill01_BJ,Liang97_BJ,Liang98_BJ}.
Briefly, we first obtain a Delaunay simplicial complex of the molecule
from weighted Delaunay triangulation, which decomposes the convex hull
of atom centers into tetrahedra (3-simplices), triangles
(2-simplices), edges (1-simplices), and vertices (0-simplices).  We
then obtain the dual simplicial complex of the protein molecule by
removing any tetrahedra, triangles, and edges whose corresponding
Voronoi vertices, edges, and planar facets are not fully contained
within the protein molecule \cite{Edels95_DCG,Edels98_DAM}.  The edges
between atoms that are not connected by bonds corresponds to nonbonded
alpha contacts.  The total sum of the number of edges for each atom is
the total number of {\it alpha contacts} $n_\alpha$. It reflects the
total number of atoms that are in physical nearest neighbor contact
with other atoms.  These atoms have volume overlap and their
corresponding weighted Voronoi cells intersect.  The {\it alpha
coordination number\/} is $z_\alpha \equiv n_\alpha/n$, where $n$ is
the total number of atoms in the molecule (see
Figure~\ref{Fig:AlphaContact}). In our calculation, we only consider
nonbonded alpha contacts. Details of the theory and computation of
alpha shape and dual simplicial complex can be found elsewhere
\cite{Edels94_ACMTG,Liang98a_Proteins,Liang98b_Proteins,LiangDill01_BJ}.

We follow previous work in reference \cite{LiangDill01_BJ} and define packing
density $p_d$ as:
\[
p_d \equiv \frac{v_{vdw}}{v_{env}}  = \frac{v_{vdw}}{v_{ms} +v_{voids}}
\]
where $v_{vdw}, v_{ms}$ and $v_{voids}$ are van der Waals volume,
molecular surface volume and the void volume of the molecule,
respectively \cite{Liang98a_Proteins}. Packing density is computed
with a solvent probe radius 1.4 \AA, as described in reference 
\cite{LiangDill01_BJ}.

{\it Growth Model for Off-Lattice Random Polymers.}  We use a modified
off-lattice discrete $m$-state model first developed in reference
\cite{ParkLevitt95_JMB} to generate self-avoiding walks (SAWs) in
three dimensional space.  All monomers are treated as balls with a
radius of 1.7 \AA.  For monomers $i$ and $j$ that are not sequence
near neighbors ($|i-j|>2$), the Euclidean distance $d(i,j)$ between
them must be greater than $2 \times 1.7$ \AA\ so they are
self-avoiding.  Sequence neighboring monomers are connected by a bond
of length 1.5 \AA.

\begin{figure}[t]
  \centerline{\epsfig{figure=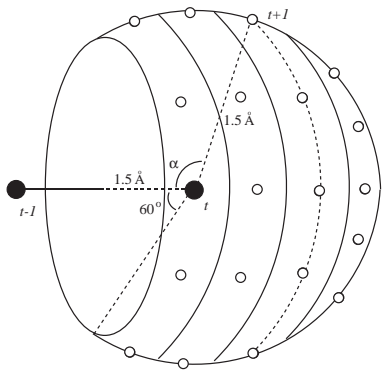,width=2.in}}
\caption{ \sf
The 32-state discrete model for chain
growth. There are 32 possible positions for adding the next
monomer. They are located on a sphere of radius 1.5 \AA, but placement
on the cap with an angle $<60$ \degree\ from the entering bond is
forbidden.  The surface is divided into four stripes of equal area.
Eigh positions are placed evenly on each strip. $\alpha$ is the bond
angle.
 }
\label{Fig:32state}
\end{figure}

We use a chain growth model to obtain conformation of polymer of
specified length \cite{Rosenbluth55_JCP}.  There are $m=32$ possible
states where the next monomer can be placed.  They are evenly
distributed spatially on a sphere of radius 1.5 \AA\ centered at the
current monomer.  We forbid the placement of the new monomer anywhere
on a cap of the sphere with an angle $< 60$ \degree\ from the entering
bond.  This ensures that there are no unnatural acute sharp bond
angles.  The remaining sphere is divided into 4 strips, each may have
different width but is of equal surface area.  For the 32 possible
states, we place uniformly 8 points at the midline of each of the 4
strips.  Following Park \& Levitt \cite{ParkLevitt95_JMB}, the
coordinates of each state are parameterized by two angels $\alpha$ and
$\tau$ for ease of computation.  $\alpha$ is the bond angle formed by
the $i-1, i$ and $i+1$-th monomers.  $\tau$ is the torsion angle
formed by four consecutive monomers.

{\it Approximately Maximum Compact Polymer.}
In addition, we generate polymers that are approximately maximum
compact based on the face centered cubic (FCC) packing of balls of
1.7\AA\ radii. For hard spheres, FCC packing has recently been
proved to have the tightest packing \cite{FCC,Sloane98_Nature}.
Because the distance between two balls in canonical FCC packing
is $2\times 1.7 = 3.4$ \AA, which is greater than the bond length
1.5 \AA, we shorten the distance along bonds connecting contacting
balls of radius 1.7\AA\ to 1.5 \AA. This mimics the bond length of
the model polymer.  Unlike FCC packing of hard spheres, bonded
monomers here are allowed to have volume overlaps.  Additionally,
there are some boundary effects because bonds connecting balls in
different layer have a distance $>1.5$ \AA. Although
mathematically unproven, we conjecture that this artificially
constructed polymer represents conformations of SAWs that have
very close to maximum compactness.

The packing density of canonical FCC packing by our method is 0.74
\cite{LiangDill01_BJ}.  As described earlier, although FCC packing
contains no voids, there are packing crevices or dead spaces that do
contribute to the calculation of $p_d$ by our definition
\cite{LiangDill01_BJ}.  In approximately maximum compact polymer,
because the distance between bonded balls is shorter than that in FCC
packing, $p_d$ can be as high as 0.80 for polymers with a range
of chain length.

\begin{figure}[t]
  \centerline{\epsfig{figure=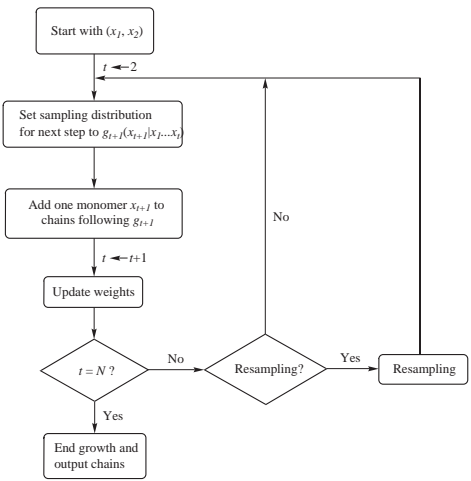,width=2.5in}}
\caption{\sf
The steps of sequential Monte
Carlo method applied to improve sampling efficiency. }
\label{Fig:flow-smc}
\end{figure}

{\it Importance Sampling with Sequential Monte Carlo.}  Since we are
simulating compact conformations that resemble proteins, we need
an efficient method to generate adequate number of conformations
satisfying protein-like compactness criteria.  Here we use a
sequential Monte Carlo (SMC) chain growth strategy
\cite{Liu&Chen98,LZC02_JCP}, which combines importance sampling and
the growth method.  The main steps are shown in
Figure~\ref{Fig:flow-smc}.

Denote the conformation of a polymer of length $t$ as $(x_1, \ldots,
x_t)$, where $x_i$ is the 3-dimensional location of the $i$-th
monomer. Starting with fixed initial location $(x_1,x_2)$, we grow
polymers by sequentially adding one monomer $x_{t+1}$ to occupy one of
the 32-states connecting to the last monomer $x_t$ of the current
chain. The monomer $x_{t+1}$ is randomly placed according to a
sampling probability $g_{t+1}(x_{t+1}|x_1\ldots x_t)$. In this study,
the following function $g_{t+1}$ is used. Let $\omega$ be one of the
32-states connected to $x_t$ that satisfies the self-avoiding
criterion.  First we intitialize the number of neighbors $n_e(\omega)$
to $\omega$ as 1, and the Euclidean distance from $\omega$ to the
nearest neighbor monomer $d(\omega)$ to 6 \AA.  We then increment
$n_e(\omega)$ by the number of existing monomers within a distance of
6.0 \AA\ to $\omega$.  Among these monomers, we identify the monomer
$x_s$ that is the nearest neighbor with the shortest Euclidean
distance $d$ to $\omega$.  We require in addition that the sequence
separation $|s-t|>3$ so $x_s$ and $x_t$ are not sequence near
neighbors. The distance $d(\omega)$ is then replaced by the value of
$d$.  The sampling probability is set as:
\[
g_{t+1}(x_{t+1}=\omega |x_1\ldots x_t) \propto 
	e^{-E'(\omega)/T'} 
\]
where $E'(\omega) =  \ln \frac{[d(\omega)]^c}{n_e(\omega)}$ is an
artificial ``packing energy'' favoring more compact conformations, and
$T'$ is a pseudo-temperature controlling the behavior of sampling.
Using this energy function, growth to position $\omega$ with close
nearest neighbor (small $d(\omega)$) and a large number of neighbors
within a 6 \AA\ distance (large $n_e(\omega)$) is favored.  Here the
adjustable parameter $c$ is used to balance the effect of $d(\omega)$
and $n_e(\omega)$.  $T'$ controls the importance of compactness.  At
low $T'$, conformations generated are compact, but at high $T'$, the
compactness criterion becomes less important.

According to the sequential Monte Carlo framework, the importance
weight $w_{t+1}$ for the sampled conformation $(x_1, \ldots, x_{t+1})$ is
updated as:
\[
w_{t+1} = w_t \cdot \frac{\pi_{t+1}(x_1\ldots x_{t+1})}
{\pi_{t}(x_1\ldots x_t) \cdot g_{t+1}(x_{t+1}|x_1\ldots x_t)}
\]
where $\pi_{t+1}(x_1\ldots x_{t+1})$ is the target distribution
at $t+1$. With a set of weighted samples
$\{(x_1^{(j)}, \ldots, x_n^{(j)}), w_n^{(j)} \}_{j=1}^m$,
statistical inference on the target distribution $\pi_n(x_1,\ldots,x_n)$
can be made using
\begin{equation}
E_{\pi_n}[h(x_1,\ldots,x_n)]=
\frac{\sum_{j=1}^m w_n^{(j)}\cdot h(x_1^{(j)}, \ldots, x_n^{(j)})}
{\sum_{j=1}^m w_n^{(j)}}
\label{Eq:expectation}
\end{equation}
for most of proper function $h$.

{\it The Target Distribution.}  We wish to generate random samples of
polymer with different compactness criterion. This is achieved by
using a target distribution $\pi_n$ which is uniform among all SAWs
satisfying a compactness constraint. The constraint is set as follows.
First, for each chosen pair values of $(T', c)$, we use the function
$e^{-E'(c)/T'}$ to generate 500 random conformations as a trial run.
Ignoring the importance weights, we calculate the mean alpha
coordination number $z_\alpha^*(n,T',c)$ of all the generated
conformations.  Then we set the target distribution $\pi_n^*$ as the
uniform distribution of all SAWs satisfying $z_\alpha\in (0.8\cdot
z_\alpha^*(n,T',c), 1.2\cdot z_\alpha^*(n,T',c))$.

We then rerun a large simulation with the same $(T', c)$
parameters and harvest the conformations, using uniform
distribution with no restriction on the intermediate target
distribution $\pi_t$ but take the truncated distribution $\pi_n^*$ as
the final target distribution. The truncation is archived by
discarding all generated conformations that does not satisfy the
constraint. Typically, the truncation rate is very small
($<0.1\%$). The bias in sampling is fully compensated by proper
weighting. 

\begin{figure}[t]
  \centerline{\epsfig{figure=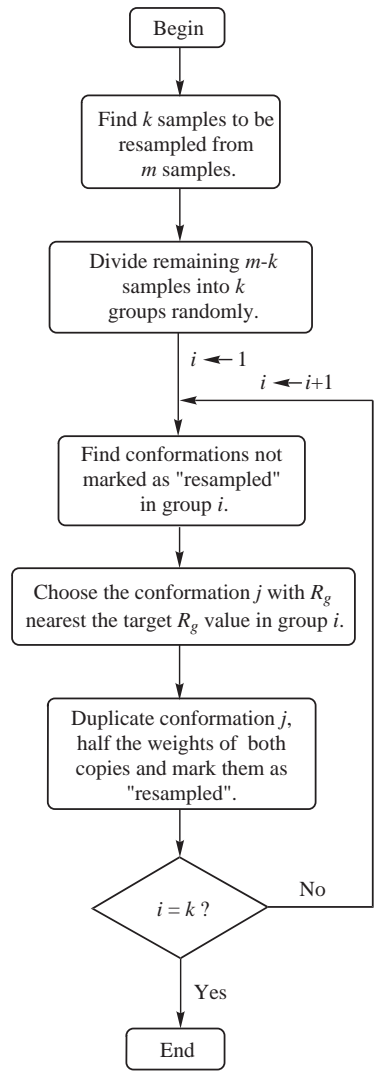,width=2.in}}
\caption{\sf
The steps of the resamplig procedure for the sequential Monte Carlo
method.
 }
\label{Fig:flow-resampling}
\end{figure}

{\it Resampling.}  Because it is easy to have self-avoiding walks to
grow into a dead-end, we use resampling to replace dead samples or
samples with small weight to improve sampling efficiency
\cite{LZC02_JCP}.  Intuitively, we check regularly during the chain
growth process whether a particular chain is stuck in a dead-end, or
is too extended, or has too little weight.  If so, this chain is
replaced by the replicate of another chain that has the desired
compactness.  Both duplicate chains will then continue to grow, and
the final two surviving chains will be correlated up to the duplication
event.  Conformations of the monomers added after the duplication will
be uncorrelated.  This resampling technique targets our simulation to
specified configuration space without introducing too much bias where
conformations all have desired compactness (see
Figure~\ref{Fig:flow-resampling}).

Although we found that the total contact number $n_\alpha$ is an
excellent parameter for characterizing protein, its calculation
involves expensive computation of weighted Delaunay triangulation and
alpha shape.  We decide to use $R_g$ as a surrogate parameter during
resampling.  For resampling, we use the empirical relationship $R_g(n)
=2.2 \cdot n^{0.38}$, where $n$ is the number of monomers in the polymer, as
described in reference \cite{Skolnick97_JMB}.  This relationship has
been used as a constraint in NMR protein structure determination
\cite{HuangPowers01_JACS}.  We have the following pseudo code for
resampling \cite{LZC02_JCP}:
\begin{tabbing}
123\=456\=789\=\kill
{\bf Procedure} {\sc Resampling} ($m, d_s, R_t$)\\
// {\sf $m$: Monte Carlo sample size, $d_s$: steps of looking-back.}\\
// {\sf $R_t$: targeting $R_g$.}\\
$k \leftarrow$  number of dead conformations.\\
Divide $m - k$ samples randomly into $k$ groups.\\
{\tt for } group $i =1$ {\tt to} $k$ \\
   \> Find conformations not picked in previous $d_s$ steps.\\
   \>  \>//{\sf Pick the best conformation $P_j$} \\
   \>  \>$P_j \leftarrow$ polymer with $\min |R_g - R_t |$\\
   \> Replace one of $k$ dead conformations with $P_j$\\
   \> Assign both copies of $P_j$ half its original weight.\\
{\tt endfor}
\end{tabbing}
Here $d_s$ is used to maintain higher diversity for resampled
conformations.  That is, conformation that has been picked in
the past $d_s$ steps are not available for resampling.

After resampling, the samples with their adjusted weights remain to be
{\it properly weighted} with respect to the original target
distribution. We can then calculate the expected alpha coordination
number $z_\alpha$, expected packing density $p_d$, or expected value
of any other function $h$ using Equation (\ref{Eq:expectation}). With these
sampling and re-sampling strategies, we can successfully grow
thousands of self-avoiding walks of chain length up to $2,000$ using a
Linux cluster of 40 CPUs.

\section{Results}

\begin{figure}[t]
  \centerline{\epsfig{figure=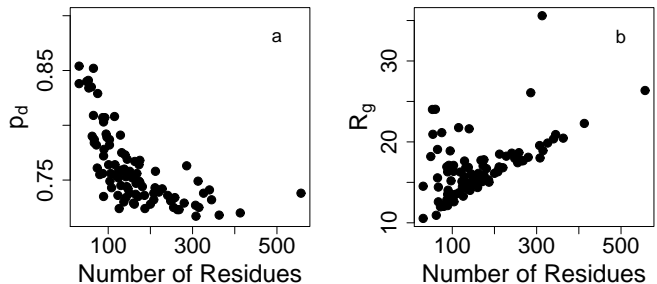,width=3.6in}}
\caption{\sf 
The relationship between packing density $p_d$, radius of
gyration $R_g$, and the number of residue $N$ in single domain
proteins.
 }
\label{Fig:1}
\end{figure}

{\it Packing Density.}  Figure~\ref{Fig:1}a shows the correlation of
packing density $p_d$ with the number of residues $N$ in real
proteins.  Similar relationship has been observed in reference
\cite{LiangDill01_BJ}.  Here we further restrict the samples to be of
single domain by {\sc Scop} annotation \cite{SCOP}. We found that
$p_d$ decreases with chain length.  That is, short chain proteins have
high packing density $p_d$, but $p_d$ decreases from $>0.85$ to about
$0.74 - 0.75$ when the chain length reaches about 190 residues.  After
reaching this length, proteins seem to be indifferent about the
existence of voids. This suggests that maintaining high packing
density is only characteristic of short chain proteins.

{\it Radius of Gyration of Proteins.}
To identify the factors that dictate the scaling behavior of $p_d$ with
residue number $N$, we need to determine whether such scaling is due
to physical constraints of statistical mechanics or the product
of extensive optimization by evolution.  We study this problem by
examining the scaling behavior of $p_d$ with $N$ in random chain
polymers generated by computer.

Because of the enormity of conformational space, we focus on random
polymers that resemble proteins in some rudimentary sense.  One
possible criterion is the radius of gyration $R_g$.  This parameter
has been widely used as a macroscopic description of protein packing.
For single domain proteins, however, we found that there is
substantial variance in $R_g$ for proteins of the same chain length
(Figure~\ref{Fig:1}b).  Therefore, $R_g$ characterizes protein packing
rather poorly, and is unsuitable as a criterion for generating
protein-like polymers for our purpose.

\begin{figure}[t]
  \centerline{\epsfig{figure=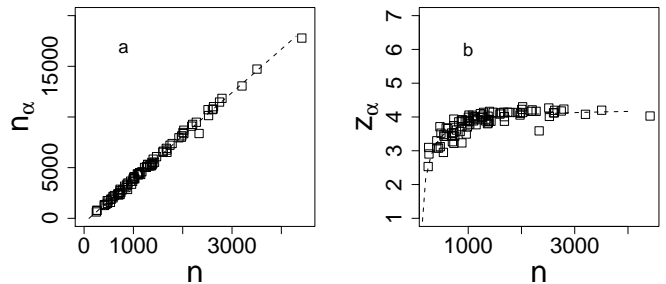,width=3.6in}}
\caption{\sf
The scaling behavior of the total number of nonbonded
atomic alpha contacts with the total number of atoms for single domain
proteins.  Here only contacts from different residues are counted.
 }
\label{Fig:2}
\end{figure}

{\it Alpha Contacts.}  An alternative global description of protein
structure is the total number of nonbonded alpha contacts $n_\alpha$
defined by the dual simplical complex of the protein.  
In Figure~\ref{Fig:2}a we plot $n_\alpha$ against the total
number of atoms in the molecule $n$.  As discussed before, these
contacts are identified by computing the dual simplicial complex of
the molecule \cite{Liang98a_Proteins,LiangDill01_BJ}.  The total
number of contacts $n_\alpha$ scales linearly with $n$.  It also
scales linearly with the protein chain length (or residue number $N$,
data not shown).  Regression leads to a linear relationship of
$n_\alpha = 4.28 \cdot n - 432$, with $R^2 = 0.995$.  The alpha
contact number $n_\alpha$ therefore provides a more accurate global
characteristic of protein than radius of gyration $R_g$.  This linear
scaling relationship of packing related property is similar to other
linear scaling relationships observed for protein, for example, of
empirical solvation energy \cite{Chiche90_PNAS}, protein surface area
and protein volume \cite{LiangDill01_BJ} with chain length.  It is
interesting to note that the value of $x$-axis intercept for $n$ of
the linear regression model suggests that the size of a minimum
protein would be in the order of 100 atoms, or about 9--10 residues.

The details of the linear scaling relationship are further examined in
Figure~\ref{Fig:2}b.  It is a replot of Figure~\ref{Fig:2}a after
normalization by $n$.  It showed that for proteins with 1,000 atoms or
more ($\ge$ 120 residues), the parameter alpha coordination number
$z_\alpha = n_\alpha/n$ is a constant of about 4.2.  
For smaller proteins ($n < 1,000$), $z_\alpha$ ranges from 2.5 to
4.0. A nonlinear curve fitting leads to the relationship $z_\alpha = a
- b/n $, where $a = 4.27 \pm 0.03$, and $b = 4.2 \times 10^2 \pm
26$. We decide to use $z_\alpha$ as the criterion to select random
polymers generated computationally for packing analysis.

\begin{figure}[t]
  \centerline{\epsfig{figure=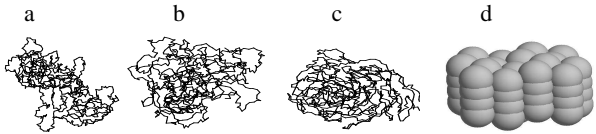,width=3.3in}}
\caption{\sf
Examples of self-avoiding walks of length 1,000 generated
with different sampling probability function $e^{-E'(c)/T'}$ using
different $(T',c)$ values.  (a).  Conformations generated with $(T',c) =
(1.0, 0.0)$; (b). $(T',c) = (0.67, 0.0)$ and (c). $(T',c) = (0.1, 0.6)$.
(d). Approximately maximally compact conformation.
 }
\label{Fig:molscript}
\end{figure}

\begin{figure}[t]
  \centerline{\epsfig{figure=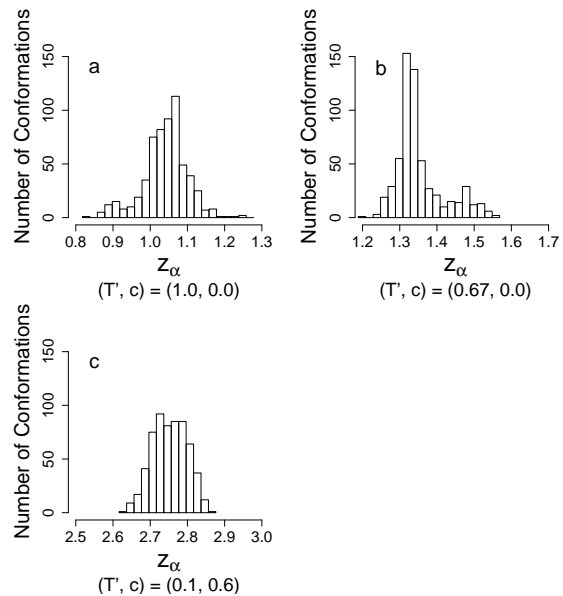,width=3.in}}
\caption{ \sf
Self-avoiding walks generated with different $(T',c)$
values do not overlap in $z_\alpha$ values.  (a).  Conformations
generated with $(T',c) = (1.0, 0.0)$; (b). $(T',c) = (0.67, 0.0)$; and
(c). $(T',c) = (0.1, 0.6)$.  Here the numbers of conformations are
unweighted. The weighted average $z_\alpha$ values are as shown in
Figure~\ref{Fig:4}a at length $n=2,000$. 
}
\label{Fig:noOverlap}
\end{figure}

{\it Targeted Sampling of Random Chain Polymer.}
Figure~\ref{Fig:molscript} shows typical conformations generated with
different $(T', c)$ parameters and the conformation of maximally
compact polymer.  Figure~\ref{Fig:noOverlap} shows the histogram of
$z_\alpha$ of the conformations at length 2,000 without weight
adjustment generated using different $(T', c)$ parameters.  It can be
seen that the histograms for different values of $(T',c)$ do not
overlap. This feature demonstrated that with properly chosen $(T',
c)$, we can efficiently generate random polymers with $z_\alpha$
within a targeted range.

\begin{figure}[t]
  \centerline{\epsfig{figure=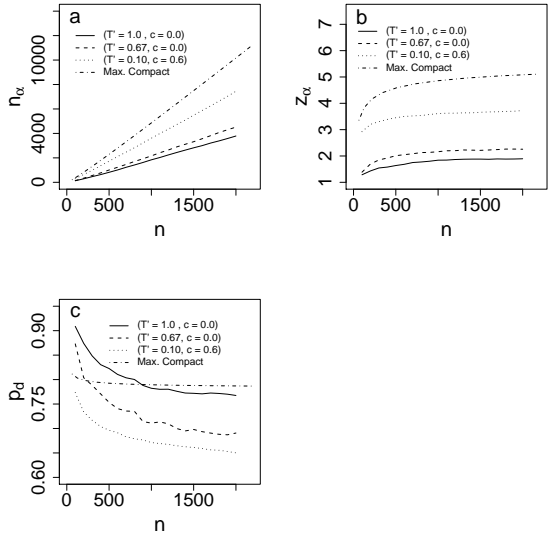,width=3.in}}
\caption{ \sf
 The relationship of alpha contact number
$n_\alpha$, alpha coordination number $z_\alpha$, and the packing
density $p_d$ of random compact and maximally compact self-avoiding
walks.  The curves are sampled following the function $e^{-E'(c)/T'}$,
with different $T'$ and $c$ values.  Each data point is an average of
10 runs of sample size 600.
}
\label{Fig:4}
\end{figure}

{\it Packing Density of Random Chain Polymer.}  Figure~\ref{Fig:4}a
shows the relationship of $c_\alpha$ associated with each pairs of
$(T',c)$ as a function of chain length $n$. It also shows the
$c_\alpha$ value for the maximally compact conformations.
Figure~\ref{Fig:4}b shows the relationship of $z_\alpha$ and $n$.
Note that the targeted $z_\alpha$ generated with different $(T', c)$
parameters give rise to different $z_\alpha \sim n$ scaling behavior.
Because the coarse grained random polymers generated here lack side
chains, they are fundamentally different from real proteins. We
therefore have experimented with several $(T', c)$ value.  We find
that protein-like scaling can be obtained for a wide range of $(T',
c)$ values.

Figure~\ref{Fig:4}b shows the average packing density $p_d$ for all
conformations satisfying the constraint specified by different
$(T',c)$ values. Except maximally compact conformations, the scaling
of $p_d \sim n$ of all other sets of polymers is remarkably similar to
that of protein (Figure~\ref{Fig:1}a).

Conformations from Set 1 ($(T', c) = (1.0, 0.0)$) are more extended,
and have lower average $z_\alpha$ ({\it e.g.},
Figure~\ref{Fig:molscript}a).  Because there are fewer voids, they
also have high $p_d$.  Conformations in Set 3 ($(T', c) = (0.1, 0.6)$)
are more compact and make more nonbonded contacts and hence have high
$z_\alpha$ values ({\it e.g.}, Figure~\ref{Fig:molscript}c). They also
form more voids, and therefore have lower $p_d$ values.  Set 2
($(T',c) = (0.67, 0.0)$) are conformations whose properties are
between those of set 1 and set 3 ({\it e.g.},
Figure~\ref{Fig:molscript}b).

\begin{table}[thb] 
 \begin{center}
\caption{ \sf
The relatinhsip between $z_\alpha$ and
$n$ can be described by the equation $z_\alpha = a - b/n$.  The
estimated values of $a$ and $b$, along with standard deviations in
parenthesis, are listed for three different sets of conformations
generated with different $(T', c)$ parameters which characterize the
sampling probability.  The values of $a$ and $b$ for real proteins are
also listed.
}
\label{curvefit}
\vspace*{.3in}
 \begin{tabular}{|c|c||c|c|}
\hline \hline
 $T'$ &$c$ & $a $ & $b $ \\ \hline \hline
1.0  & 0.0 & 1.88(0.02) & 72(7) \\ \hline
0.67 & 0.0 & 2.24(0.02) & 67(6) \\ \hline
0.1  & 0.6 & 3.68(0.02) & 93(7) \\ \hline \hline
\multicolumn{2}{|c||}{Native Proteins} & 4.27(0.04) & 423(27) \\ \hline 
\hline
 \end{tabular}
 \end{center}
\end{table}

The relationship between $z_\alpha$ and chain length $n$ can be
characterized by a nonlinear equation $z_\alpha = a - b/n$, similar to
that of real proteins.  The sets of $a$ and $b$ obtained by curve
fitting  are listed in Table~\ref{curvefit}. We
emphasize that these randomly generated self-avoiding walks are
fundamentally very different from proteins: all residues are of
uniform size, there are no side chains, and there is no hydrophobic or
any other type of physical interactins in these polymers.  Because it
is impossible to quantitatively define a similarity metric that
measures how different these polymers are from proteins, we are not
able to decide which specific values of $(T', c)$ are optimal for
modeling protein packing.  Nevertheless, the scaling of $p_d \sim n$
for all $(T', c)$ values is qualitatively quite similar to that of
real proteins.

\begin{figure}[t]
  \centerline{\epsfig{figure=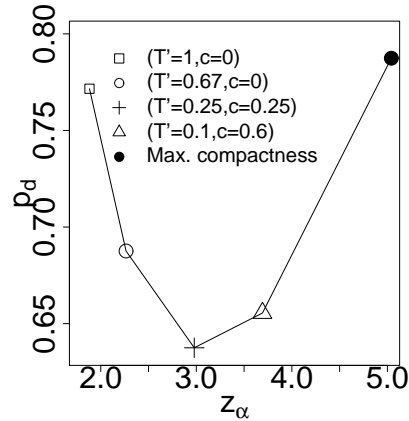,width=2.5in}}
\caption{\sf
The relationship of $p_d$ and $z_\alpha = n_c/n$  for self-avoiding walks
generated using different $(T', c)$ parameters at chain length 1,800.
 }
\label{Fig:Ucurve}
\end{figure}

The relationship between $p_d$ and $z_\alpha$ at chain length 1,800
for self avoiding walks generated with different parameters $(T', c)$
are shown in Figure~\ref{Fig:Ucurve}.  The $p_d$ of both extended and
maximally compact conformations have high $p_d$ values, but
conformations with an intermediate value of $z_\alpha$ contain voids
and have smaller $p_d$ values.  This is similar to the relationship of
$p_d$ and a compactness parameter $\rho$ (equivalent to $z_\alpha$
used here) studied in two dimensional lattice (see Figure 8 in
reference \cite{LZC02_JCP}).

\section{Summary and Discussion}
It is well acknowledged that protein has high packing density $p_d$,
as high as that of crystalline solids \cite{Richards94_QRB}.  However,
recent study suggested that there are numerous voids and pockets in
proteins \cite{LiangDill01_BJ}.  It was also found that about 1/3 of
the residues in a protein deviates from the FCC close packing and have
random positions \cite{Bagci02_JCP}.  The simulation results presented
here indicate that chain connectivity, excluded volume, and global
compactness are the main determinants of the scaling behavior of voids
and chain length in proteins.  Unlike maximally compact polymers which
maintains high packing density at all chain length, proteins and
simple near compact polymers have large $p_d$ values only for
relatively short chains.  When the chain length reaches 190 residues
for protein and about 600--700 for chain polymers, proteins and
polymers have lower packing density and are quite tolerant to the
formation of voids.

The global compactness is a necessary condition for the observed
protein-like scaling behavior.  However, not all parameters related to
voids and compactness are equally appropriate.  The data shown in
Fig~\ref{Fig:2} suggests that the alpha coordination number $z_\alpha$
reflects basic intrinsic compactness properties of protein, which is
absent in the widely used parameter $R_g$, the radius of gyration.  The
advantage of parameters such as $z_\alpha$ emphasizes the importance
of accurate description of protein geometry and structure.

The parameters $p_d$ is biased towards short chain proteins.  By
definition, a polymer formed by 2, 3 or a small number of monomers do
not have long enough chains to form voids, therefore all will have
$p_d = 1.0$.  When chain length becomes longer, voids appear.
Similarly, $z_\alpha$ is also biased towards short chains.  For very
short chain polymers where the chain has few turns, few non-bonded
contacts exist and no voids are formed.  In this case, $z_\alpha$ is
low and $p_d$ is high.  However, this small size effect disappears
rapidly for our model conformations for maximally compact polymers.
Small size effect therefore does not fully account for the scaling
behavior of $p_d$ and $z_\alpha$ in proteins and in simulated random
polymers.

There are major differences between self-avoiding walks we generated
and real protein structures.  Our SAWs have no side chains, and belong
to the coarse-grain model where one monomer is represented as a ball.
In addition, the target distribution of SMC sampling is the truncated
uniform distribution of all geometrically feasible conformations.  The
truncation required is that polymers must satisfy a prescribed
$z_\alpha \sim n$ relationship.  No physical forces such as
hydrophobic interactions is used in the target distribution.

In this paper, we describe a novel approach to overcome the attrition
problem in generating long chain compact self-avoiding walk.  With
sequential importance sampling and resampling, we have developed an
algorithm that effectively sample rare events, {\it i.e.}, compact
self-avoiding walks.  Success in generating thousands of off-lattice
self-avoiding walks satisfying various desired compactness requirement
is essential for studying the scaling behavior of $p_d$ in random
off-lattice self-avoiding walks.

The main result of this paper is that with sequential Monte Carlo
techniques it is not difficult to reproduce protein-like scaling
behavior of packing density $p_d$ and chain length $n$ in generic
chain polymers.  With the guidance of rudimentary requirement of
$z_\alpha$, this can be achieved under a wide range of $z_\alpha \sim
n$ relationships.  We therefore conclude that proteins retain the same
packing property of generic compact chain polymers.  We further
conclude that proteins are unlikely to be optimized by evolution to
eliminated packing voids.  This is in support of the insightful
comments of Richards who suggested that an appropriate level of
under-packing would be important for evolution to occur through random
mutations \cite{Richards97_CMLS}.

Our study showed the importance of generic geometric packing related
to $z_\alpha$ in reproducing protein-like $p_d \sim n$ scaling
behavior.  To test further the role of geometric packing, the next
step would be to examine the $p_d \sim n$ scaling by generating more
realistic compact random polymers with perhaps monomers of different
sizes to model the side chain effects.  Furthermore, with sequential
Monte Carlo and other advanced sampling methods, various models of
explicit side chains can be attached to main chain monomers.  In
addition, one could introduce various alphabet sets for the residues
(such as the HP model) and corresponding potential energy function $H$.
In this case, the target distribution can be the Boltzmann distribution
$\pi \propto \exp (-H/T)$ instead of the uniform distribution of all
SAWs.  It would also be interesting to examine the $p_d \sim n$
scaling of polymers of random sequences and of protein-like sequences
with low energy in compact states.

\section{Acknowledgments}
We thank Profs.\ Herbert Edelsbrunner and Luhua Lai for valuable
discussions. This work is supported by funding from National Science
Foundation CAREER DBI0133856, DBI0078270, DMS0073601, CCR9980599
and American Chemical
Society/Petroleum Research Fund 35616-G7.

\end{document}